\documentclass[fleqn,10pt]{wlscirep}
\usepackage{amsmath}
\usepackage{amssymb}
\usepackage{graphicx}

\title{Nondestructive in-line sub-picomolar detection of magnetic nanoparticles in flowing complex fluids}

\author[1,+,*]{Lykourgos Bougas}
\author[2,+]{Lukas D. Langenegger}
\author[2]{Carlos A. Mora}
\author[2]{Martin Zeltner}
\author[2]{Wendelin J. Stark}
\author[1]{Arne Wickenbrock}
\author[3]{John W. Blanchard}
\author[1,3,4,5]{Dmitry Budker}
\affil[1]{Johannes Gutenberg-Universit{\"a}t  Mainz, 55128 Mainz, Germany}
\affil[2]{Functional Materials Laboratory, Department of Chemistry and Applied Biosciences, ETH Zurich, CH-8093 Zurich, Switzerland}
\affil[3]{Helmholtz-Institut Mainz, 55128 Mainz, Germany}
\affil[4]{Department of Physics, University of California, Berkeley, CA 94720-7300, USA}
\affil[5]{Nuclear Science Division, Lawrence Berkeley National Laboratory, Berkeley, CA 94720, USA}
\affil[*]{lybougas@uni-mainz.de}
\affil[+]{these authors contributed equally to this work}

\begin{abstract}
Over the last decades, the use of magnetic nanoparticles in research and commercial applications has increased dramatically. However, direct detection of trace quantities remains a challenge in terms of equipment cost, operating conditions and data acquisition times, especially in flowing conditions within complex media. Here we present the in-line, non-destructive detection of magnetic nanoparticles using high performance atomic magnetometers at ambient conditions in flowing media. We achieve sub-picomolar sensitivities measuring $\sim$30\,nm ferromagnetic iron and cobalt nanoparticles that are suitable for biomedical and industrial applications, under flowing conditions in water and whole blood. Additionally, we demonstrate real-time surveillance of the magnetic separation of nanoparticles from water and whole blood. Overall our system has the merit of inline direct measurement of trace quantities of ferromagnetic nanoparticles with so far unreached sensitivities and could be applied in the biomedical field (diagnostics and therapeutics) but also in the industrial sector.
\end{abstract}

\begin{document}
\flushbottom
\maketitle
\thispagestyle{empty}

\section*{Introduction}
Functionalized magnetic nanoparticles have emerged as unique objects for a variety of applications, ranging from usage in life sciences\,\cite{Lu2007} to data storage\,\cite{Reiss2005} and industrial wastewater treatment\,\cite{Xu20121}. Especially, an increasing number of applications in biomedicine, such as in diagnostics, imaging, drug delivery and other therapeutic approaches, use magnetic nanoparticles due to their intrinsic properties\,\cite{Rotello2004}. In any of those applications the magnetic properties of the nanoparticles are either used for their detection and subsequent readout of information (e.g. diagnostic devices), or for physical manipulation of the particles mostly in separation processes. \\
\indent Health and environmental risks associated with the release of and exposure to engineered nanoparticles are unclear and hamper many applications especially in the medical field. Those risks could be either addressed by the utilization of inherently safe nanoparticles or by the prevention of exposure to the nanoparticles. While the assessment of the safety of nanomaterials advanced over the last years and standards are being established, residual risks remain, as biological interactions are complex and effects could occur years after exposure. Therefore, a promising alternative is to prevent any exposure by implementing means of sensitive detection combined with shutdown mechanisms that seal contaminated parts.\\
\indent One concept, which would particularly benefit from the implementation of such a measurement as part of the safety mechanism, is magnetic particle-based blood purification (MPBP). MPBP has been proposed as a new therapeutic approach to remove disease causing factors such as toxins, proteins, or whole pathogens directly from a patient's blood in an extracorporeal circuit (see Ref.\,\cite{Frodsham2015} and references therein). Magnetic nanoparticles functionalized with capturing moieties (e.g. antibodies) are applied in an extracorporeal circuit where they bind to their targets and are separated by a magnetic field before the blood is recirculated to the patient. Several setups utilizing this concept have been proposed and tested \textit{in vitro}\,\cite{Weber1997,Stamopoulos2007,Yung2008,Herrmann2010,Herrmann2015}, as well as in animal trials\,\cite{Herrmann2013,Kang2014}.\\
\indent Translation from a lab concept to clinical applications in humans demands thorough assessment of possible risks for the patients. Although some ferromagnetic particles have proven to be bio-compatible\,\cite{Stamopoulos2010,Herrmann2011a,Bircher2014,Jacobson2015a,Herrmann2016aa}, the best way to minimize acute and long term exposure risks is to fully remove the particles from blood before it is recirculated to the patient\,\cite{Schumacher2013}. Detection of small amounts of nanoparticles in flowing blood coupled to a shutdown mechanism that stops the procedure in case of a failure could, therefore, drastically reduce overall risks associated with MPBP\,\cite{Herrmann2015}. Up to date, particle detection methods applied in the context of MPBP have been inadequately sensitive [$\geqslant$\,1\,ppm (w/w)]\,\cite{Herrmann2015}, but most importantly indirect and destructive\,\cite{Herrmann2013,Kang2014,Schumacher2013}, and thus, impractical for implementation in a therapeutic setting. In general, detection of trace quantities of nanomaterials is very challenging due to low sensitivities and serious matrix effects. In addition, the often very expensive instrumentation, destructive nature of measurement, tedious sample preparation, and long processing times, prevent real-time analysis of small nanoparticle concentrations, which are highly desired.\\
\indent Metallic ferromagnetic nanoparticles are particularly beneficial for separation processes in complex media, as they exhibit higher saturation magnetizations and allow for fast and complete separation using high gradient magnetic separators. In addition, the detection of the magnetic moment of metallic ferromagnetic nanoparticles is promising and such a direct method could be applied in most media without matrix effects avoiding sample preparation or destruction. This can be achieved using magnetometric methods, which are not based on optical data acquisition and, therefore, can operate in complex media, such as opaque whole blood. However, the magnetic moments of nanoparticles are extremely small, and thus, any employed measurement methods need to be correspondingly sensitive.\\ 
\indent There exist several magnetic-particle detection technologies that have demonstrated competitive sensitivity relevant to the aforementioned applications: superconducting quantum interference devices (SQUIDs)\,\cite{Chemla2000,Horng2005,Fong2005}, giant magneto-resistive (GMR) sensors\,\cite{Pannetier1648,Osterfeld2008,Wang2014}, atomic magnetometers\,\cite{Kominis2003,Shah2007}, and diamond-based magnetometers\,\cite{Gould2014,Glenn2015,Plakhotnik2015}. The requirements in terms of size, price, operating conditions, usability, and need for sample preparation, restricts broad application of many of those technologies, especially to MPBP-related applications. Atomic-based magnetometers though, offer highest magnetic field sensitivity, allow for non-invasive and non-destructive sensing modalities, and can be operated at ambient conditions (no cryogens) requiring only magnetic shielding. Atomic magnetometers have been recently used for the detection of magnetic micro-particles in flow conditions\,\cite{Xu2006,Maser2011,Yu2012}.\\
\indent In this work, we establish a setup and data analysis method that employs a high performance atomic magnetometer for the real-time, in-line and non-destructive, detection of ferromagnetic nano-particles in flowing water and whole blood. We first evaluate the detection sensitivity and establish a particle concentration-signal relationship for nano-particle solutions. The system is then applied for the surveillance of the separation of magnetic particles from flowing water and whole blood. For both matrix media, especially in whole blood, we achieve particle detection sensitivities in the sub-picomolar range, which is at least two-orders-of-magnitude more sensitive than previous, indirect methods employed in MPBP.\\
\section*{Materials and Methods}
\subsection*{Magnetic particles} 
We use iron-carbide- and cobalt-based magnetic nanoparticles, as well as superparamagnetic nanoparticles, to demonstrate the sensitivity and applicability of our in-line magnetometric detection system. In particular, for our measurements we use the following particles:
\begin{itemize}\setlength\itemsep{-2pt}
\item[$\circ$] \textbf{C/Fe3C:} Carbon-coated iron carbide particles, mean diameter $\sim$24\,nm, approximate density 7.9\,g/ml, and saturation magnetization $\sim$79\,emu/g (see supplementary materials for characterization data).
\item[$\circ$] \textbf{C/Co:} Carbon-coated cobalt particles, mean diameter $\sim$34\,nm, approximate density 8.9\,g/ml, and saturation magnetization $\sim$118\,emu/g (see supplementary materials for characterization data).
\item[$\circ$] \textbf{Nanomag\textsuperscript{\textregistered}-D -} Micromod's 09-00-132 Nanomag\textsuperscript{\textregistered}-D dextran coated superparamagnetic iron oxide particles, mean diameter $\sim$130\,nm, and saturation magnetization $\sim$45\,emu/g. 
\end{itemize}
\indent We choose these particular cobalt- and iron-carbide-based ferromagnetic particles because they show larger saturation magnetizations compared to oxide-based particles, and their characteristics and size are suitable for MPBP-related applications\,\cite{Herrmann2010,Herrmann2013}. We note here that, particle characterisation and magnetization measurements suggest that the bulk number of particles is ferromagnetic and that contamination with superparamagnetic particles - if any at all - are below 0.1\% (see supplementary materials). In addition, the ferromagnetic particles are coated with a hydrophilic polymer that that reduces surface fouling and agglomeration and thereby permits the preparation of stable dispersions of these nanoparticles in water and in biological fluids. Furthermore the polymer allows for further attachment of (bio)chemical molecules\,\cite{Zeltner2012,Hofer2015,Schumacher2013}.\\
\indent The nanomag\textsuperscript{\textregistered}-D particles are acquired commercially\,\cite{micromod} and we use them as a reference for our measurements in water solutions. In particular, since the detection takes place within a magnetically shielded environment (i.e. under zero external magnetic field), we expect that any (super-) paramagnetic system will not produce a detectable signal as long as the relaxation properties of the particles remain unchanged and are not affected by the matrix media\,\cite{Weaver2012,Chemla19122000}.\\
\indent We note here that characterisation data for the particles we have used in our measurements can be found in the supplementary materials. \\
\subsection*{Experimental setup}
\begin{figure}[t]
\centering
\includegraphics[width=\linewidth]{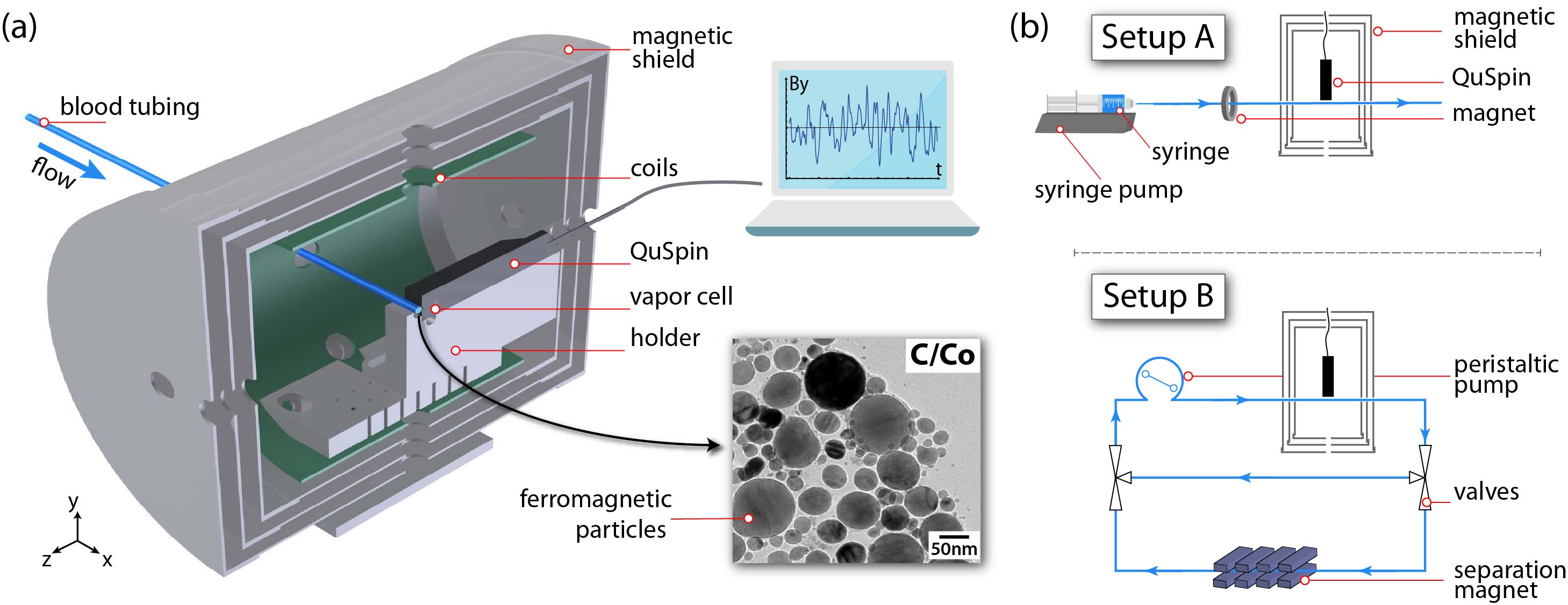}
\caption{\small{(a) Schematic diagram of the in-line magnetic sensor used for ferromagnetic particle detection inside a magnetic shield (QuSpin is the atomic magnetometer). (b) Diagrams for the different setups used in this work: Setup (A) uses a syringe pump for fluid displacement. Particles are additionally pre-magnetized with a ring magnet before passing the sensor. In setup (B) solvents are circulated using a peristaltic pump. The fluid can be circulated either in a closed loop without magnetization, or, by switching the three-way valves, through a high-gradient magnet that allows for the magnetic particle separation.}}
\label{fig:Fig1}
\end{figure}

The primary motivation for this work is the design of an in-line, non-destructive, magnetic sensor employed in a extra-corporeal circuit for MPBP using functionalized polymer-coated, water-dispersible ferromagnetic nanoparticles. The experimental setup employed in this work is, thus, designed and optimized to simulate the conditions under which an atomic-based magnetic sensor is incorporated into an MPBP extra-corporeal circuit, as schematically presented in Fig.\,\ref{fig:Fig1}. \\ 
\textbf{Magnetometric sensor.} The magnetic sensor is a spin-exchange relaxation-free (SERF)\,\cite{Kominis2003} optically pumped atomic magnetometer (OPAM) (QuSpin Inc.\,\cite{Quspin}), consisting of a $^{87}$Rb vapor cell with dimensions of 3$\times$3$\times$3\,mm$^3$, which defines the sensing volume. OPAM operating in the SERF regime, are the most sensitive magnetometers operating in the low-frequency regime\,\cite{Allred2002,Bang2010}. A feedback system holds the temperature of the vapor cell at the levels required for operation in the SERF regime, and thus any temperature changes induced by the flow of the solution are compensated and do not affect the function of the magnetometer, while the thermally isolating housing design allows for outside surface temperatures no more than a few degrees higher than ambient temperature. The magnetometer is located in the center of a cylindrical magnetic shield which provides sufficient suppression of ambient external magnetic fields. We measure the magnetic sensitivity of our OPAM to be 16.9\,fT/$\sqrt{\rm{Hz}}$ along the y-axis and 16.7\,fT/$\sqrt{\rm{Hz}}$ along the z-axis (averaged from 1\,Hz to 100\,Hz) with a bandwidth of $\sim$145\,Hz (see supplementary materials)\,\cite{Savukov2017}.\\
\textbf{Setups.} We employ two setups (A) (linear) and (B) (recirculating) [Fig.\,\ref{fig:Fig1}]. In both systems we use medical-grade blood-infusion tubing lines (Fresenius Heidelberger Extension Lines with an inner and outer diameter of 3.0\,mm and 4.1\,mm, respectively) to continuously flow solutions of magnetic particles in close proximity to the OPAM. The distance from the center of the vapor cell of the OPAM to the outer side of the magnetometer's protective jacket is typically 6.4\,mm. For the QuSpin magnetometer employed in setup (A), an additional custom slit allows us to place the outer surface of the tubing closer to the cell by an additional 1.5\,mm. In each setup, the tubing is held in position on top of the magnetometer [and along the slit in setup (A)], and the whole system (magnetometer and tubing) is secured within the center of a magnetic shield to minimize vibrations. \\
\indent In setup (A), fluid is pumped with a syringe pump (New Era Pump Systems NE-1010 High Pressure Single Syringe Pump) and for magnetic shielding we use a Twinleaf MS-2 magnetic shield (four-layer with end caps, 46\,mm axial hole, shielding factor 10$^6$, specified noise floor $\sim$25\,fT/$\sqrt{\rm{Hz}}$)\,\cite{Twinleaf}. Moreover, the tubing goes through a ring magnet (outer diameter 50.8\,mm; inner diameter 25.4\,mm; strength at the center $\sim$\,100\,mT), which is located at a distance of $\sim$40\,cm before the magnetometer, and allows for additional magnetization of the particles. The positioning and the strength of the magnet are such that no particle separation is taking place while the solutions flow through the system. \\
\indent In setup (B) we use a peristaltic pump (Lead Fluid BT100L) for continuous circulation through a tubing circuit. We use a ZG-206 magnetic shield (Magnetic Shield Corp.; 3 layer with end-caps, 22\,mm axial hole, shielding factor $\sim$1500), and three-way valves to allow for the redirection of the fluid through a high gradient magnetic filter used for the nanoparticle separation\,\cite{Herrmann2011}.\\
\textbf{Data acquisition.} Finally, for data acquisition we use the digital and analog outputs provided by the control software and hardware of the QuSpin magnetometer. The digital output allows for single-axis acquisition operation and data acquisition rates of 200\,samples/sec, resulting in a Nyquist frequency of 100\,Hz, while using the analog output we are able to perform dual-axis acquisition and obtain sampling rates of 50\,ksamples/s/ch using a data acquisition card (NI-9239; 24-bit resolution). We note here, that the magnetometer we use has a 6th order low-pass filter on the analog output, limiting the bandwidth to 145\,Hz (see supplementary materials). For our measurements we typically use the internal data acquisition software module provided by QuSpin, but we also verify that our measurements are in agreement with the results obtained via the analog outputs. \\
\subsection*{Experimental Procedures}
We first measure the magnetic field produced by different nanoparticles in water and blood dispersions at constant flow [setup (A)]. We prepare aqueous and blood solutions of varying concentrations of magnetic particles by mixing 20\,ml of distilled water or whole blood (bovine whole blood, acid citrate dextrose anticoagulation, Fiebig-N\"ahrstoffetechnik) with different amounts of volumes from a pre-sonicated starting water:particle solution of 1\,part-per-thousand in concentration, using air-displacement pipettes [we use inductively coupled plasma mass spectrometry to independently verify the concentration of our water solutions; see supplementary materials]. For the water solutions we sonicate and then vortex mix the solutions for $\sim$10\,min. For the blood solutions, we solely vortex-mix the solutions for $\sim$3\,min. Subsequently, we insert the particle solutions in a permanent-magnet Halbach-array system for several minutes to magnetize the particles (the magnetic field strength of this Halbach array system is $\sim$2\,T, which is close to the saturation field of the particles; see supplementary materials). Following this pre-magnetization step, we again sonicate and vortex mix the solutions (or solely vortex mix for the blood solutions), and then introduce them into syringes (Omnifix\textsuperscript{\textregistered} 50\,ml, B-Braun). Using the syringe pump, we are able to circulate the solutions through the tubing system bringing them in close proximity to the OPAM, and perform the magnetometric measurements. We note here that the additional ring-magnet [setup A] is used as a pre-magnetization-measurement stage, but it is not sufficient to fully saturate the magnetization of the particles.\\
\indent Using the setup (B) (Fig.\,\ref{fig:Fig1}), we measure the real-time magnetic separation of particles in water and whole blood (porcine whole blood, slaughter byproduct\,\cite{Cooper2013}, oxalate anticoagulation, provided by the Veterinary Service City of Zurich) dispersions. For these measurements, we prepare our samples from an aqueous starting solution by dilution with distilled water or porcine blood, respectively. Again, we sonicate and vortex mix the aqueous samples, while we solely vortex mix the blood samples after particle addition. For the measurements, we fill the whole circuit with the particle dispersions, which are then circulated through the tubing for a few seconds before we direct the flow through the high-gradient magnetic separator followed by the magnetometric sensor.
\section*{Results and Discussions}
\begin{figure}[h!]
\centering
\includegraphics[width=0.9\linewidth]{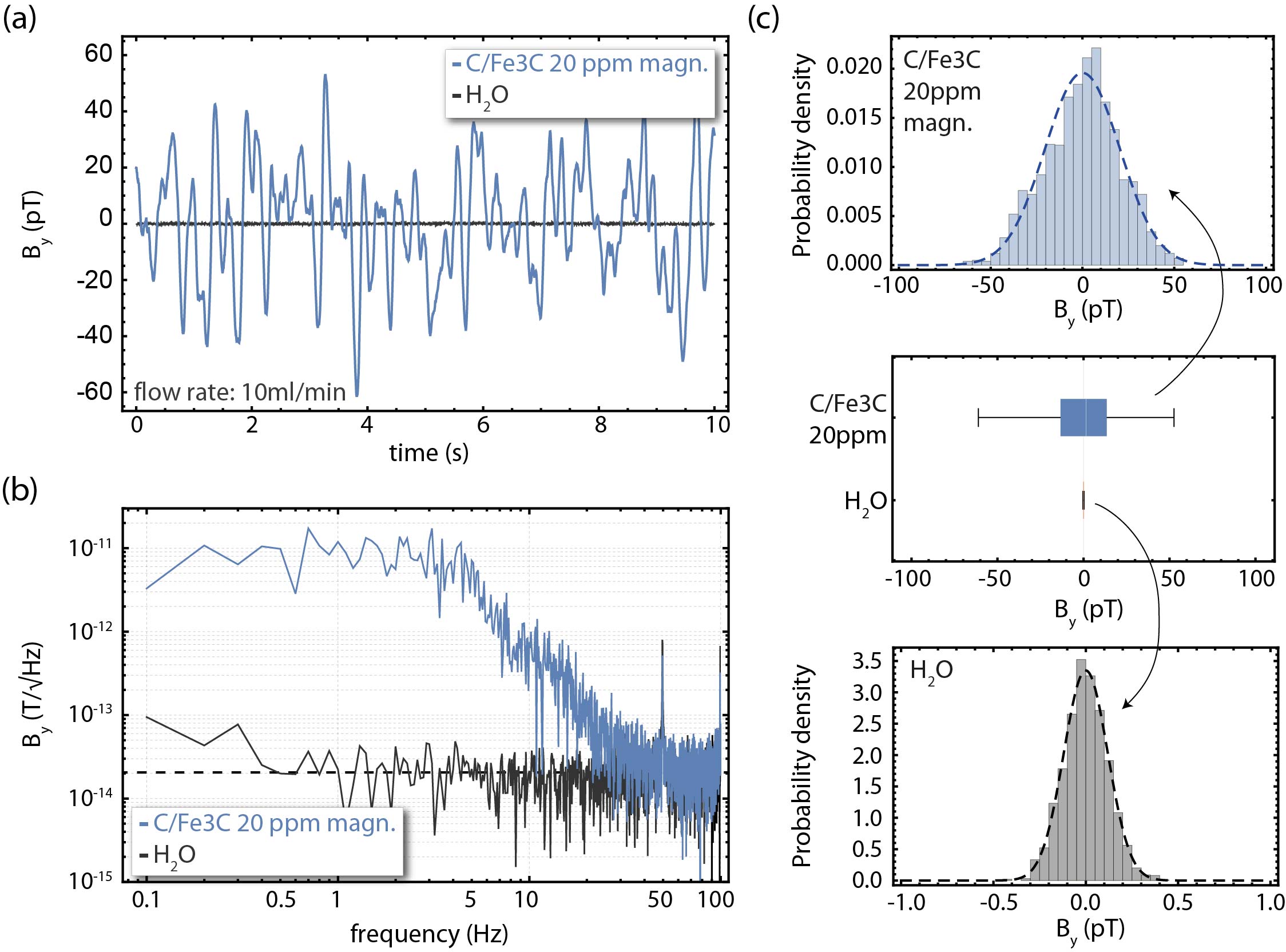}
\caption{\small{Detection of C/Fe3C ferromagnetic particles in water solutions. (a) Real time magnetic-field measurements along the y-axis for pure water (black points) and 20\,ppm:water solution of pre-magnetized C/Fe3C nanoparticles (blue points). (b) Magnetic field power spectra for the measurements presented in (a). The black dashed line corresponds to 20.4\,fT/$\sqrt{\rm{Hz}}$ (averaged from 1-100\,Hz), indicating the noise floor baseline of the measurement and verifying that the presence of the water does not influence significantly the magnetometer's sensitivity. (c) Histogram analysis of the measurements shown in (a) and (b), including a box-and-whisker plot analysis for visual comparison of the observed magnetic-field distributions (the white line represents the median marker, the grey boxes the upper and lower 25\% quantiles, while the grey bars represent the maximum and minimum acquired data points, including data outliers). For the histogram analysis we employ a digital filter to remove the 50\,Hz and 100\,Hz power-line-related noise components. All measurements are performed under constant flow conditions, with a flow rate of 10\,ml/min.}}
\label{fig:tz579water}
\end{figure}

\indent In Fig.\,\ref{fig:tz579water}\,(a) we present real-time magnetic field measurements (along the y-axis) for 10\,s, for a 20\,ppm:water solution of C/Fe3C nanoparticles flowing in a constant rate of 10\,ml/min (flow along the x-axis) through the system [setup (A)]. As a baseline for our measurements we record the response of the magnetometer while pure water flows through our system [Fig.\,\ref{fig:tz579water}\,(a)]. When magnetic particles are present, we observe large magnetic field fluctuations with a wide frequency spectral decomposition [Fig.\,\ref{fig:tz579water}\,(b)].  Using a statistical analysis algorithm we estimate the goodness-of-fit of the observed signals to a normal distribution, from which we obtain the variance of the measured magnetic field signal, which is a measure of the magnetic fluctuations, and which we show is directly related to the concentration of magnetic particles. In Fig.\,\ref{fig:tz579water}\,(c) we present the histograms for these real-time measurements along with a box-and-whisker-plot comparison between the signals obtained for pure water and for the 20\,ppm C/Fe3C:water solution. We note that for all the statistical analyses in this work we use a digital filter that removes the power-line-related noise frequency components (50\,Hz and 100\,Hz). Moreover, the OPAM has a flat response over the broad frequency-spectrum components present within the detected signal [Fig.\,\ref{fig:tz579water}\,(b); see supplementary materials], ensuring the validity of our analysis without additional calibrations. We choose the 10\,ml/min flow rate for our measurements to allow for a sufficient number of magnetic field oscillation within the selected 10\,s time window for better statistical analysis, while maintaining long acquisition times. We note here that for the chosen flow rate of 10\,ml/min, the residence time of the fluid in the sensing region, which is approximately 0.2\,ml (see discussion in the supplementary materials) is $\sim$1.2\,s. We also ensure that faster flow rates do not affect the detected signal (see supplementary materials).\\ 
\begin{figure}[h!]
\centering
\includegraphics[width=\linewidth]{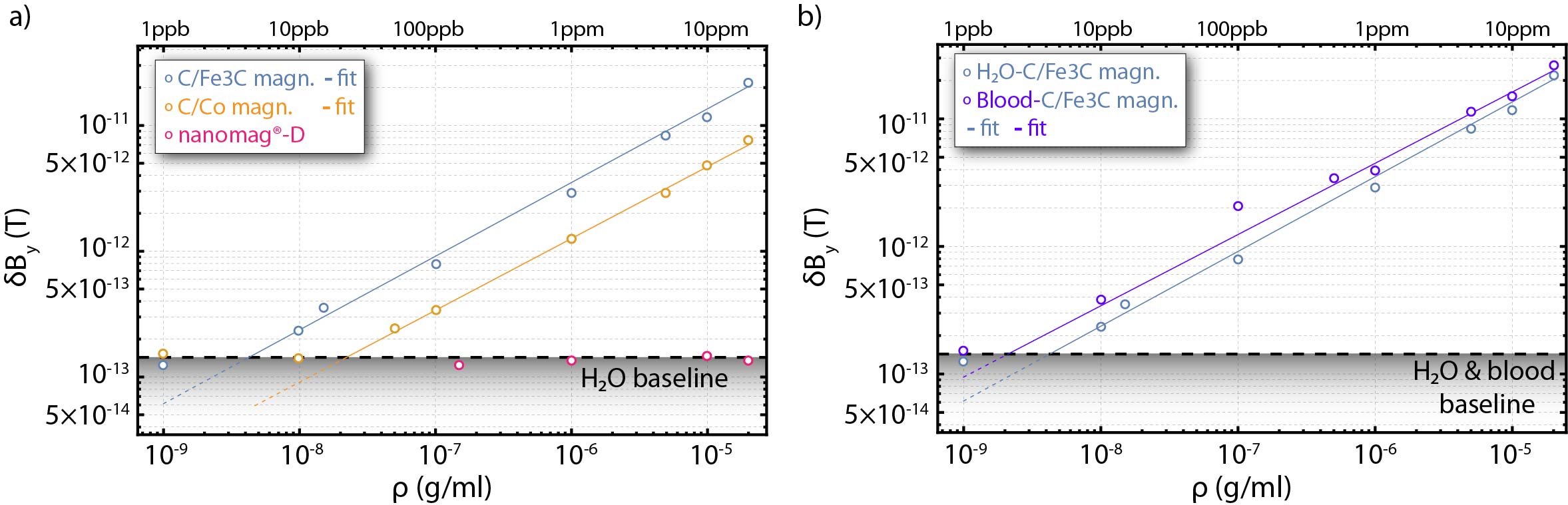}
\caption{\small{(a) Magnetic field variance signal for different concentrations of the C/Fe3C, C/Co and nanomag\textsuperscript{\textregistered}-D nanoparticles, under pre-magnetized conditions, in water solutions. (b) Comparison of the magnetic field variance signal between water and blood solutions for different concentrations of the C/Fe3C nanoparticles under pre-magnetized conditions. All measurements are realized under constant flow conditions (10\,ml/min). The black dashed lines [(a), (b)] represent the baseline for each measurement, obtained for measurements performed with pure water or blood. The solid lines represent fits to the experimental data.}}
\label{fig:TZvsSPHR}
\end{figure}
\indent We repeat the same procedure and analysis for different concentrations of C/Fe3C:water and C/Co:water solutions. Once again, for each solution and measurement, we analyze the raw time-series magnetic field measurements in consecutive 10\,s segments, and the square-root of the magnetic variance of the detected signals is estimated. We present our results in Fig.\,\ref{fig:TZvsSPHR}\,(a). For both particles, we observe a proportionality between the square-root of the magnetic-field variance and the concentration, following approximately a square-root dependence to the particle concentration ($\delta$B$\propto \rho^{0.58}$; see supplementary materials for further discussions). This is expected for random fluctuations of independent magnetic nanoparticles. Using the magnetometric measurement obtained for pure water as the measurement baseline, we can estimate the sensitivity of our sensing protocol at the $\sim$5\,ppb level for the C/Fe3C nanoparticles, and at the $\sim$30\,ppb level for the C/Co particles. Thus, we obtain $\sim$150\,fM and $\sim$280\,fM particle-concentration sensitivities for the C/Fe3C and C/Co nanoparticles, respectively (we consider here their respective composition and size distribution - see supplementary materials; in terms of detection sensitivities in iron and cobalt, we demonstrate 60\,nM and  420\,nM iron and cobalt sensitivities, respectively). Furthermore, we note here that considering the magnetic-field range in which OPAM, operating in the SERF regime, demonstrate maximum sensitivity and operate (up to $\sim$10\,nT)\,\cite{Allred2002}, we obtain that our sensing protocol has in principle a dynamic range of more than seven orders of magnitude. Finally, in Fig.\,\ref{fig:TZvsSPHR}\,(a) we present the results for our measurements using nanomag\textsuperscript{\textregistered}-D:water solutions of relatively high concentrations. As expected for superparamagnetic particles, we do not observe a magnetic signal. \\ 
\indent In Fig.\,\ref{fig:TZvsSPHR}\,(b) we compare measurements of C/Fe3C:water solutions with measurements of C/Fe3C:blood solutions. We observe larger magnetic field fluctuations for the C/Fe3C:blood particle solutions compared to the water solutions, but a similar functional dependence of the magnetic-signal variance to the particle-concentration. Since the measurement baseline is similar for blood as for pure water, the larger signals result in improved particle-sensitivity limits, yielding detection limits for C/Fe3C particles in blood-solutions of approximately 2\,ppb, which corresponds to $\sim$60\,fM C/Fe3C-particle-concentration sensitivities and 60\,nM in iron-concentration sensitivities. \\
\indent We note here that we have performed additional characterisation measurements to identify the hydrodynamic radius of the particles used in this work in water and blood solutions. In particular, we perform dynamic light scattering measurements\,\cite{Bantz2014} that suggest that the hydrodynamic radius of the particles remains approximately the same in water and in blood solutions, and is $\sim$250-300\,nm, excluding thus the possibility that the signal increase is the result of particle agglomeration (see supplementary materials). The actual mechanism of this effect is currently not fully understood. Although we can assume that this is a result of the matrix medium, the actual mechanisms for this increase in signals will be the subject of future studies. \\
\indent It is important to emphasize here that the sensitivity of the presented measurement may significantly improve further by the use of magnetic-flux guides/concentrators\,\cite{Kim2016aa}, where, for example, the detection of single micron-size particles has been recently demonstrated using a similar detection system\,\cite{Kim2017}.\\
\indent As a final demonstration, we use setup (B) to show the in-line real-time magnetic-separation detection using two different matrix media; water and porcine blood\,\cite{Cooper2013}. The flow-rate in this system is set to 14\,ml/min, therefore, a full cycle-time for the 20\,ml circuit volume is approximately 86\,s. We note here that for the flow rate used in our measurements, we do not observe any hemolysis after centrifugation of the blood samples. We can estimate the pressure drop in our flowing circuit to be $\sim$1.8\,mmHg using the Darcy-Weisbach equation for laminar flows. Compared to a physiological blood pressure of 120\,mmHg this is unproblematic and, thus, we ensure that the chosen flow rate is safe for blood handling.\\
\indent For the analysis of the magnetic separation we use the same procedure as described above (see Fig.\,\ref{fig:tz579water}). The raw time series data are analyzed in consecutive 10\,s segments, and a statistical analysis algorithm estimates the goodness of fit to a normal distribution that additionally yields the variance of the recorded magnetic field signal distribution within the analyzed time window. We present the results in Fig.\,\ref{fig:blutwasser}. The observed large magnetic field deviations during the acquisition for both matrix media can originate from particle agglomerates (a bubble chamber is also used to ensure that there are no bubbles in the flow system). While these statistical outliers yield broader magnetic distributions during the measurement analysis, we clearly observe magnetic-particle separation as an overall decrease of the magnetic-variance signal with time, reaching eventually the device baseline. Considering the baseline for each measurement (obtained from a measurement with only pure water or blood), we can infer our detection sensitivity in terms of particle concentration. For the water separation measurements [Fig.\,\ref{fig:blutwasser}\,(a)], we achieve nanoparticle separation at the level of $\sim$95\,ppb residual concentration after approximately 3 full-circuit passes (corresponding to approximately 0.5$\times10^9$ particles/ml), while for the porcine-blood separation measurements [Fig.\,\ref{fig:blutwasser}\,(b)], we achieve a nanoparticle separation at the level of $\sim$220\,ppb residual concentration after approximately 8 full-circuit passes (corresponding to approximately 1.2$\times10^9$ particles/ml). We note here that in the blood separation measurements we observe larger signal fluctuations compared to the water separation measurements, in accordance with our calibration measurements [Fig.\,\ref{fig:TZvsSPHR}\,(b)]. We note here, that possible particle agglomerates, while these might reduce the effective particle surface available for binding target substances in MPBP applications, should be easier to separate magnetically. We finally note here that the presented statistical analysis methodology is not unique, but for the purposes of this work and for the demonstration of the simultaneous separation and detection is sufficient.\\
\begin{figure}[h!]
\centering
\includegraphics[width=\linewidth]{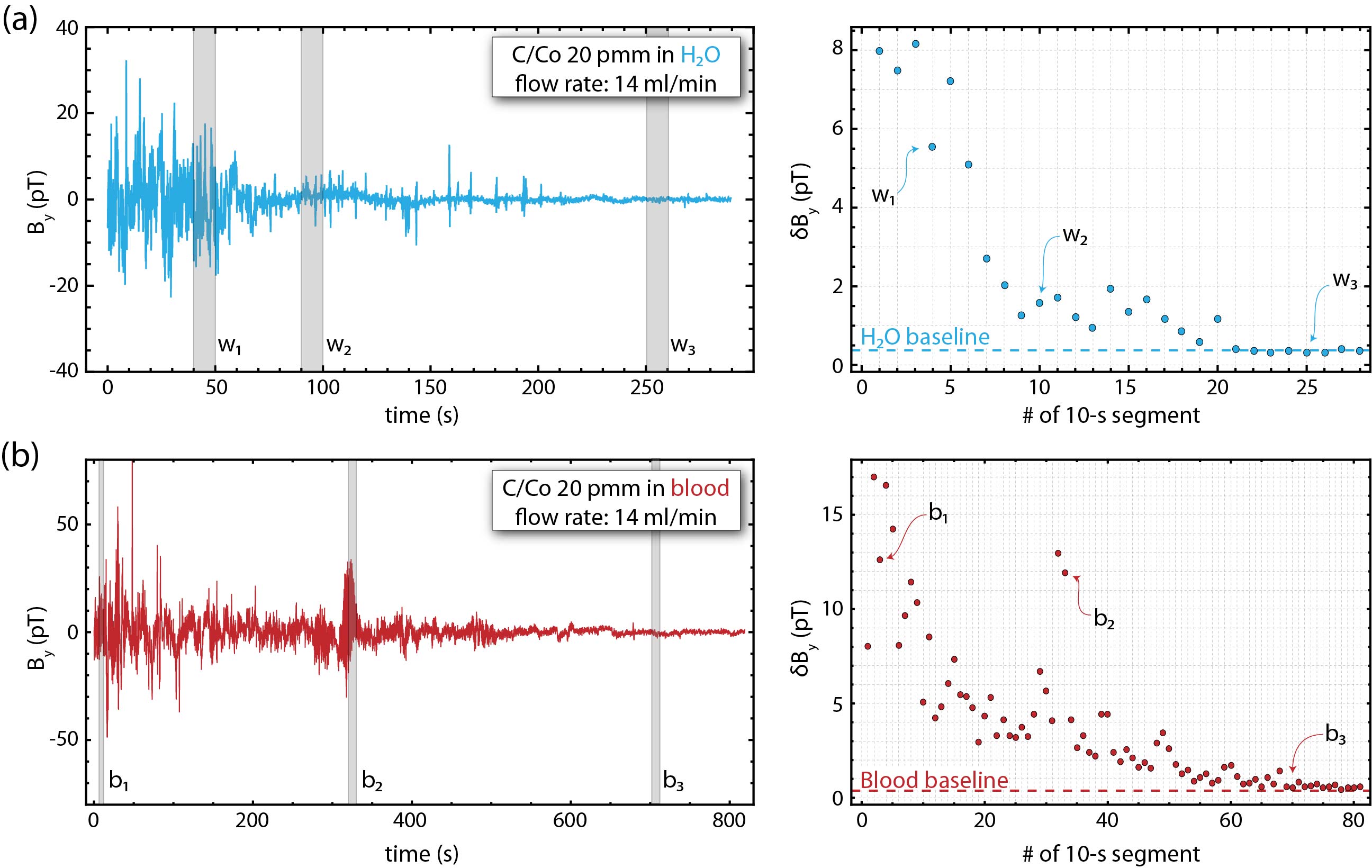}
\caption{\small{Real-time magnetic field measurements demonstrating nanoparticle magnetic separation for two different matrix media: (a) water (upper right and left plots); (b) blood (lower right and left plots). For the measurements we use 20\,ppm C/Co:water/blood solutions. The large magnetic fluctuations observed during the separation measurements in blood [b2 in (b)] are attributed to particle agglomeration (see supplementary materials).}}
\label{fig:blutwasser}
\end{figure}
\section*{Conclusions}
We have demonstrated a highly sensitive, in-line, non-destructive, magnetic nanoparticle sensor based on a high performance atomic magnetometer, which operates at ambient conditions and reaches sub-picomolar nanoparticle sensitivities. This sensitive in-line measurement of particles could greatly benefit applications where remaining magnetic nanoparticles could impose a risk to humans and the environment if released from the respective systems. \\
\indent The primary motivation for our work is the in-line measurement of remaining (ferro-)magnetic particles used in extracorporeal blood purification, a treatment modality that enables rapid removal of poorly accessible high molecular weight disease-causing compounds from blood. While previous demonstrations have been destructive and of moderate sensitivity, using high performance atomic magnetometers we demonstrate sub-picomolar particle detection limits in a non-invasive, and thus non-destructive, modality. Additionally, the presented sensing modality and detection approach has broad applications in other research and industrial fields. Prominent example are in-line measurement of remaining (ferro-)magnetic particles used to adsorb and extract heavy metals and other harmful substances from waste water\,\cite{Xu20121}, and the in-line detection of wear-generated magnetic debris from transmissions and gearboxes measured in circulating hydraulic oil and other lubrication fluids as early warning method for machine failure\,\cite{oilwaste}.\\

\bibliography{report}
\section*{Acknowledgements}
LB acknowledges support by a Marie Curie Individual Fellowship within the second Horizon 2020 Work Programme, and thanks P. Bl{\"u}mler and L.\,N. Cohen for fruitful discussions. DB and AW acknowledge support by the German Federal Ministry of Education and Research (BMBF) within the Quantumtechnologien program (FKZ 13N14439). JWB acknowledges support from the Helmholtz Postdoc Programme. We are grateful to D. K\"aser for her help with characterisation measurements, and in particular to Dr. V.\,K. Shah and Dr. O. Alem from QuSpin Inc. for their constant support and help.  


\section*{Author contributions}
L.\,L., C.\,A.\,M., L.\,B., and D.\,B. conceived the experiments. L.\,B., L.\,L., and C.\,A.\,M. conducted the experiments. L.\,B. analyzed the results. M.\,Z. developed the ferromagnetic particles. A.\,W., J.\,W.\,B., W.\,J.\,S., and D.\,B. supervised the project. All authors discussed the results and contributed to the writing of the manuscript.

\section*{Additional information}
L.\,L., C.\,M., and W.\,J.\,S. are co-founders of hemotune Ltd. and have financial interests in the company. The remaining authors declare no competing financial interests. 

\section*{Data Availability:} The datasets generated during and/or analyzed during the current study are available from the corresponding author on reasonable request.

\end{document}